\documentclass[%
 reprint,
 amsmath,amssymb,
 aps,
]{revtex4-2}

\usepackage{graphicx}
\usepackage{dcolumn}
\usepackage{xcolor}
\usepackage{caption}
\usepackage{subcaption}
\usepackage{bm}
\usepackage{array}
\newcolumntype{P}[1]{>{\centering\arraybackslash}p{#1}}
\usepackage{xcolor}

\begin{document}

\preprint{APS/123-QED}

\title{Emergent dynamics and spatiotemporal patterns in soft robotic swarms}

\author{R. Pramanik$^{1,2}$, R.W.C.P. Verstappen$^1$ and P.R. Onck$^{2,}$}
\email{Corresponding author: p.r.onck@rug.nl}

\affiliation{$^1$Computational \& Numerical Mathematics Group, Bernoulli Institute for Mathematics, Computer Science \& Artificial Intelligence, University of Groningen, Netherlands\\
$^2$Micromechanics Group, Zernike Institute for Advanced Materials, University of Groningen, Netherlands}%


\date{\today}

\begin{abstract}
The collective swimming of soft robots in an infinite viscous fluid is an emergent phenomenon due to the non-reciprocal hydrodynamic interactions between individual swimmers. These physical interactions give rise to unique spatiotemporal patterns and unusual swimming trajectories that are often difficult to predict due to the two-way fully coupled nature of the strong fluid-structure interaction at a thermodynamic state that is far from equilibrium. Until now, robotic swarms have mostly been studied for rigid swimmers in two-dimensional settings. Here we study the emergence  of three-dimensional spatiotemporal patterns of helical magnetically actuated  soft-robotic  swimmers by systematically studying the effect of different initial configurations. Our results show that swimmers with variations in initial positions in the swimming direction are attracted to each other, while swimmers with variations in lateral positions repel each other, eventually converging to a state in which all swimmers concentrate in one lateral plane drifting radially outward.
\end{abstract}


\maketitle


\section{\label{sec:level1}INTRODUCTION}

The study of active matter has gained significant attention in recent years due to its diverse applications across biological systems, discrete materials, and synthetic structures \cite{nguyen2018engineered,shaebani2020computational,ning2024macroscopic}. This includes granular matter, bacterial swarms, flocking behaviors, and stimuli-responsive gels, all of which exhibit complex, far-from-equilibrium self-organization driven by hydrodynamic interactions \cite{zuo2020dynamic, tian2021collective, wu2015collective}. Within this context, systems such as magnetotactic bacteria and swimming cells demonstrate non-reciprocal fluid-mediated interactions that lead to emergent behaviors like vortex formation and viscous merging \cite{gardi2023demand, kokot2018manipulation, lushi2014fluid}. These phenomena are not only theoretically intriguing but also have practical implications in the development of advanced microrobotic systems \cite{hagan2016emergent}.

Focusing on the burgeoning field of magnetically actuated microrobotics, there is a growing interest in using magnetic fields to control and actuate micro- and nanoscale robotic swarms for biomedical applications. These systems offer promising solutions for targeted drug delivery, minimally invasive surgeries, and parallelized manipulation in medical settings \cite{yang2021motion, chen2015occlusion}. Recent advances have demonstrated various locomotion modes - such as rolling, flipping, and corkscrew motions - enabled by magnetic actuation, which are particularly useful in navigating complex biological environments \cite{zimmermann2022multimodal, dong2019collective}. Among these, rotating magnetic field-driven microrobots have shown great potential, especially in applications requiring precise control and localization. For instance, kinematic models have been developed to predict the behavior of magnetic screws in soft tissues, and reconfigurable collective modes of magnetically actuated disks have been explored, paving the way for innovative biomedical interventions \cite{chaluvadi2020kinematic, basualdo2022control}.

Theoretical models have been proposed to study the collective behavior of miniaturized swarms \cite{fujiwara2014self}; e.g., control of emergent robotic systems has been reported using Smooth Particle Hydrodynamic (SPH) models \cite{pac2007control}. The collective motion of magnetic spinners and rollers has been studied using mathematical models for the swarm particles that were subjected to magnetic torques based on Maxwell's equations \cite{wang2019quantifying}. The hydrodynamic states were reported to capture the two-dimensional collective behavior of self-assembled spinners at the water-air interface \cite{kokot2015emergence}. Another study on robot-robot interactions has been proposed to predict the evolution of swarm macroscopic properties (such as separation, compressibility, and cohesivity) based on SPH control of robotic swarms \cite{mitikiri2021smoothed}.

\begin{figure*}[htpb!]
    \centering
    \includegraphics[scale=0.4]{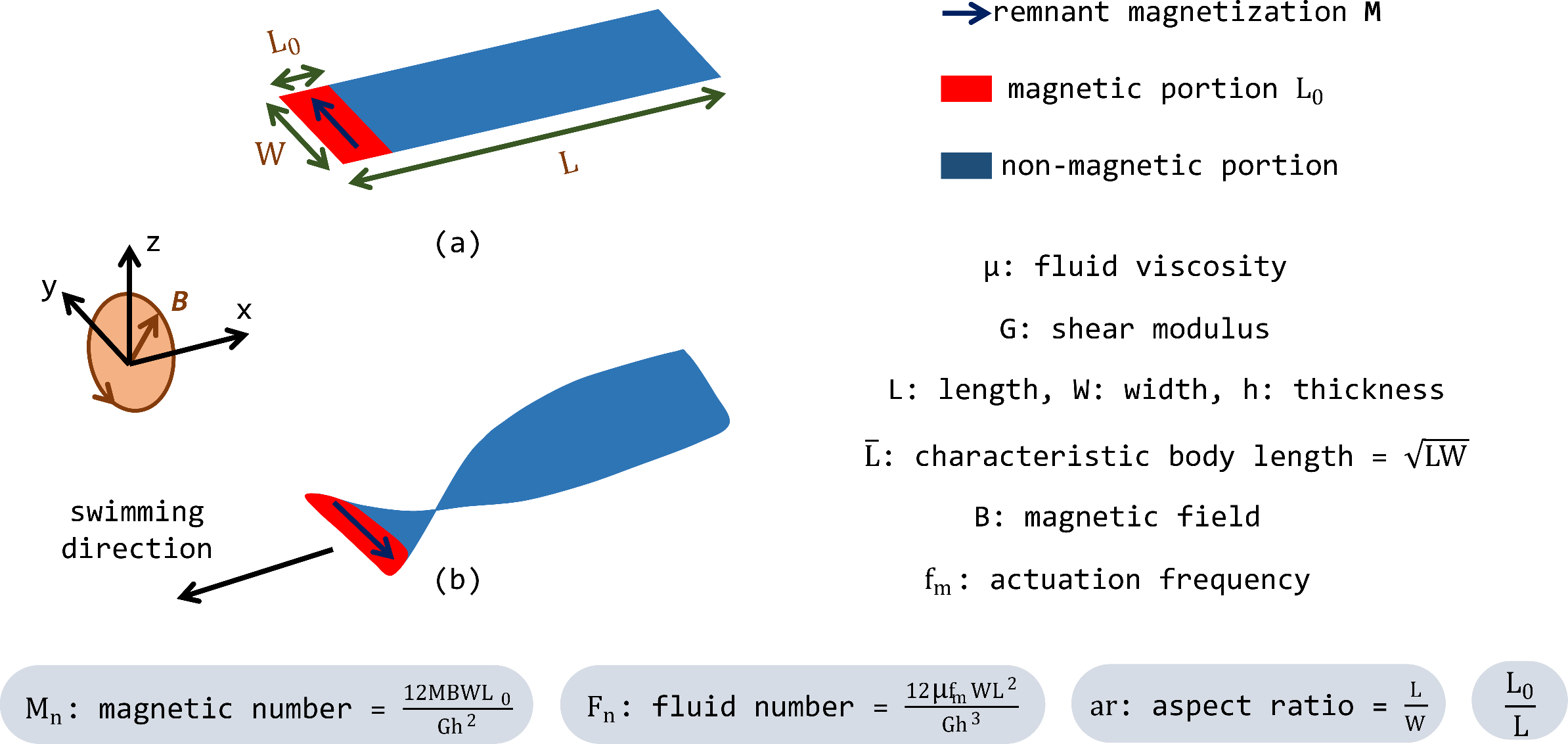}
    \caption{(a) Schematic illustration of a partially magnetic soft robotic swimmer exposed to an external magnetic field. (b) Propulsion is generated by applying an external uniform rotating magnetic field B with the rotation axis aligned along the x-axis. Chirality is naturally formed on-the-fly as a result of the interplay between viscous drag forces acting on the entire swimmer and magnetic body torques concentrated on one end of the swimmer (multimedia available online).}
    \label{fig:helicalswimmer}
\end{figure*}

Recently, spinning magnetic micron-sized disks has been shown to self-organize at the air-water interface under rotating external magnetic fields to generate various spatiotemporal patterns based on an appropriately defined interaction potential \cite{wang2022order}; here, the local pairwise interactions (between two disks) induced global patterns. In addition, the collective spatial behavior was shown to depend on the external actuation frequency based on patterns produced from Monte Carlo simulations \cite{xie2019programmable}. Swarms of ferromagnetic microparticle ensembles at the air-water interface were subjected to a uniaxial oscillating magnetic field that lead to self-organization and clustering \cite{kokot2017dynamic}. While the particle dynamics was described using a lattice spring model (where a network of Hookean springs connect point masses), the two-dimensional fluid dynamics as well as the fluid-structure interactions were captured by the lattice Boltzmann model \cite{ashurst1976microscopic,succi2001lattice,alexeev2005modeling}.

Although praiseworthy achievements have been made in the field of microrobotic swarms, these have only been reported for two-dimensional in-plane (hydrodynamic) systems of rigid swimmers, such as spheres \cite{liang2021magnetic}, rollers \cite{han2024manipulation}, spinners \cite{wang2019quantifying}, particles \cite{martinez2018emergent}, disks \cite{ehrenstein2023magnetic} and colloids \cite{wang2018reconfigurable}. However, it is important to note that these rigid robots suffer large setbacks with system miniaturization, resulting in reduced swimming modalities that hinder their applicability in biomedical applications. Additionally, precise path control, maneuverability, and localization of these rigid magnetic swarms is quite difficult. To overcome this, magnetic soft robotic swimmers have been developed, demonstrating improved swimming techniques, reliable path control, precise localization, and adaptive swimming behavior \cite{hu2018small,ren2021soft,wang2020untethered,yerecent,zhou2021magnetically,quashie2022magnetic,wang2024multi,pramanik2023magnetic}. However, the operation of these inherently soft, magnetically deformable swimmers under three-dimensional swarming conditions has not been explored before. Here, we study the emergent behavior and collective swimming of miniaturized magnetically actuated soft robotic swimmers \cite{pramanik2024nature,pramanik2024computationalexperimentaldesignfast}, where the individual swimmers communicate with each another through long-range hydrodynamic interactions in a three-dimensional setting. We will study how different initial spatial configurations, specifically in-plane and out-of-plane arrangements, affect the emergent behavior and collective swimming dynamics. For this we use a fully coupled computational model that integrates fluid dynamics, solid mechanics, large deformation solid-fluid interaction, and magnetics in a unified framework \cite{khaderi2012fluid}.

The paper is structured as follows: we begin by validating the numerical convergence of our model for configurations involving one and two helical swimmers. We then present findings on the emergent behavior of two swimmers with different initial configurations in a three-dimensional setting. Additionally, we examine the spatiotemporal patterning and collective swimming dynamics of three helical swimmers, with particular attention to their relative spacing as a function of cycle number. We discuss how the aspect ratio of the swimmers influences their collective swimming and corresponding trajectories and we close with a summary and conclusions.

\section{\label{sec:level2}RESULTS \& DISCUSSION}

For the present study, we consider partially magnetized elastica that are membrane-like and shaped in the form of rectangular flaps with only one end magnetized [see Fig. \ref{fig:helicalswimmer} (multimedia available online)] \cite{namdeo2014numerical}. By applying an external rotating magnetic field, magnetic torques are applied at the magnetic end, while the passive (i.e., non-magnetic) portion is subjected to drag forces from the surrounding fluid. Therefore, as a natural consequence of fluid-structure interaction in combination with the localized magnetic body torque, these swimmers develop a chirality (twisted body profile) and propel through the fluid using a typical corkscrew motion. Due to their miniaturized length scales, the Reynolds number of the flow is approximately equal to zero (Stokes flow). We use a robust fully coupled computational model (that simultaneously accounts for solid mechanics, fluid dynamics, large deformation fluid-structure interaction, and magnetics in a monolithic manner \cite{khaderi2012fluid}) to study the emerging behaviour of magnetically actuated helical soft robotic swimmer collectives involving two-way fluid-structure interactions as well as hydrodynamic effects. For all our simulations, we consider the magnetic (M$_n$) and fluid (F$_n$) numbers to have values of 50 and 5, respectively (refer \cite{namdeo2014numerical} for details); the default value of f$_m$ is chosen as 5Hz.

\subsection{Numerical convergence: 1 swimmer}

\begin{figure}[htpb!]
    \centering
    \includegraphics[scale=0.46]{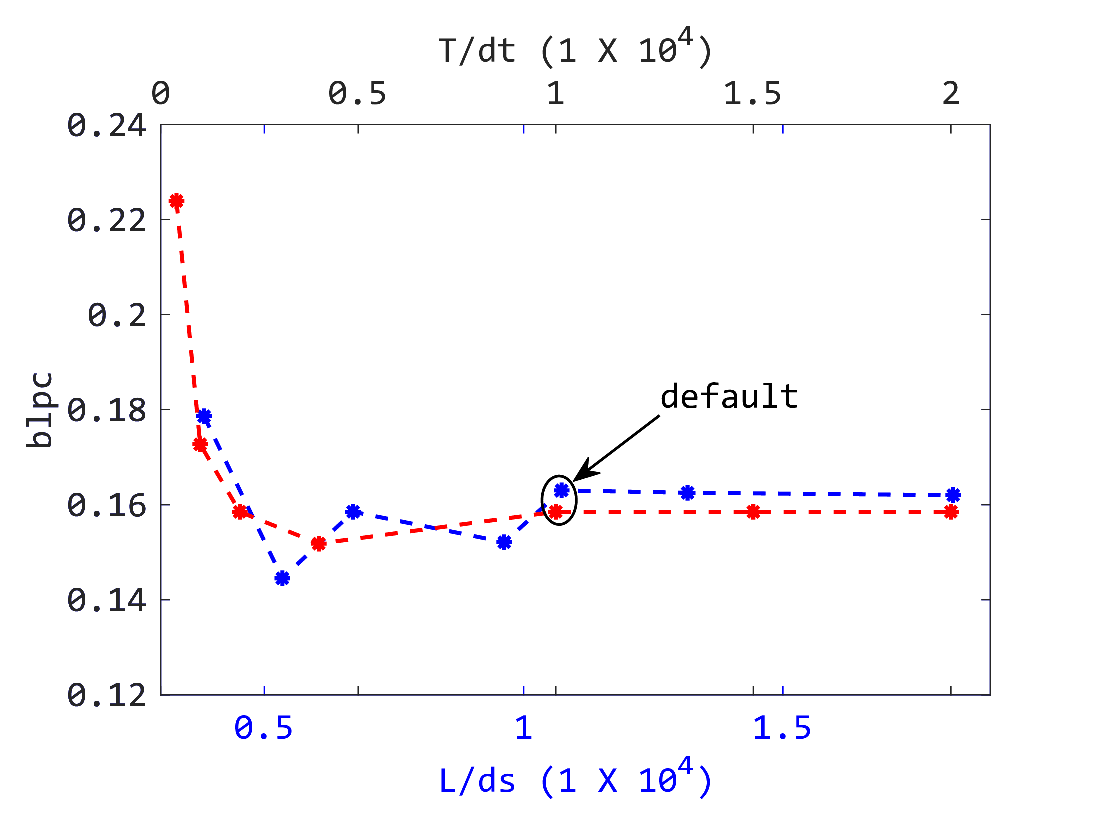}
    \caption{Convergence study and the default values corresponding to dt (in red) and ds (in blue) chosen for further simulations henceforth.}
    \label{fig:convergence}
\end{figure}

We begin with the numerical convergence study to ensure the robustness and stability of the computational framework, and also fix our default values of time step (dt) and mesh size (ds). For the convergence test, a magnetically actuated helical soft robotic swimmer is subjected to rotating magnetic fields to achieve a steady-state swimming speed reported in body lengths per cycle, blpc=c/$\mathrm{\Bar{L}}$f$_m$ with c the average swimming speed, $\mathrm{\Bar{L}}$ the characteristic body length, and f$_m$ the actuation frequency. We plot the spatial and temporal convergence results in Fig. \ref{fig:convergence} with T=1/f$_m$. We observe that while the former has an oscillatory convergence behavior, the latter has a monotonic convergence. Furthermore, the default values of ds and dt have been denoted, and these are used for all simulations henceforth.

\subsection{Numerical convergence: 2 swimmers}

\begin{figure}
    \centering
    \includegraphics[scale=0.56]{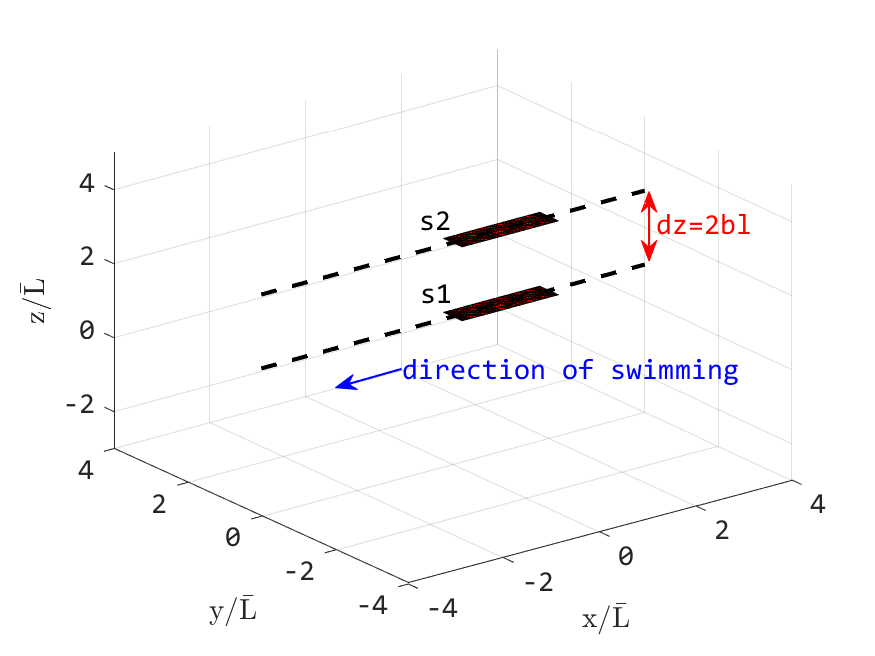}
    \caption{Schematic representation of initial (spatial) configurations of two soft robotic swimmers chosen for numerical convergence of the effect of hydrodynamic interactions for different values of ds and dt.}
    \label{fig:snapshotnumerics}
\end{figure}

We now report the swimming trajectories of 2 swimmers that are initially spaced two bodylengths (bl) apart along the z-axis; their initial configuration is schematically shown in Fig. \ref{fig:snapshotnumerics} with swimmer 1 (s1) at (y,z) = (0,1) and swimmer 2 (s2) at (y,z) = (0,3). Remarkably, while these swimmers propel along the negative x-axis using helical propulsion (and maintaining a steady shape profile) due to the application of (constant in magnitude and) rotating magnetic field in the y-z plane, we observe that they additionally start revolving around each other due to hydrodynamic interactions (multimedia available online).

\begin{figure}
    \centering
    \includegraphics[scale=0.56]{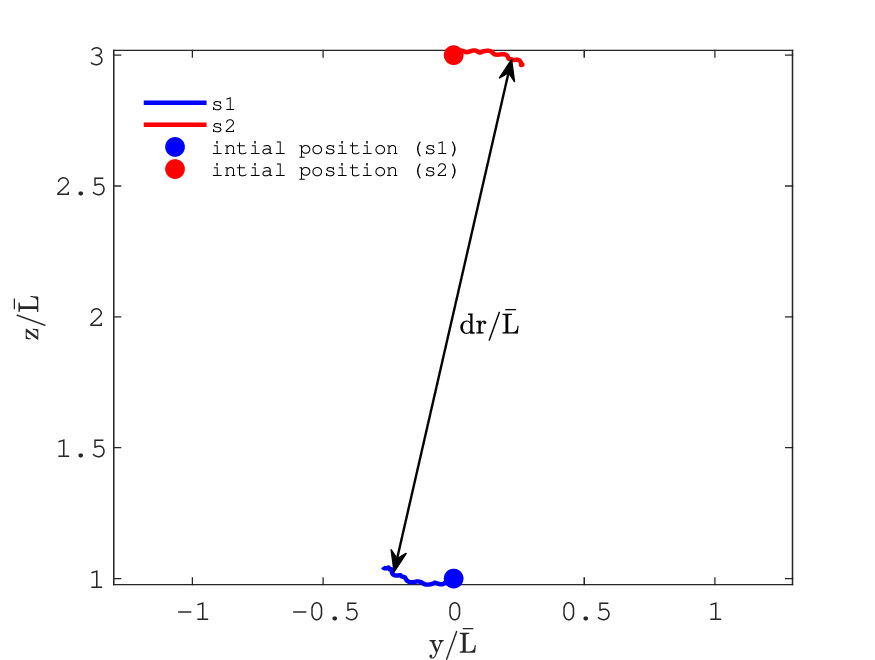}
    \caption{Phase plot of the two swimmers with initial condition as Fig. \ref{fig:snapshotnumerics} that shows their swimming trajectories and starting positions.}
    \label{fig:numerics2s}
\end{figure}

\begin{figure*}
     \centering
     \begin{subfigure}{0.49\textwidth}
         \centering
         \includegraphics[scale=0.56]{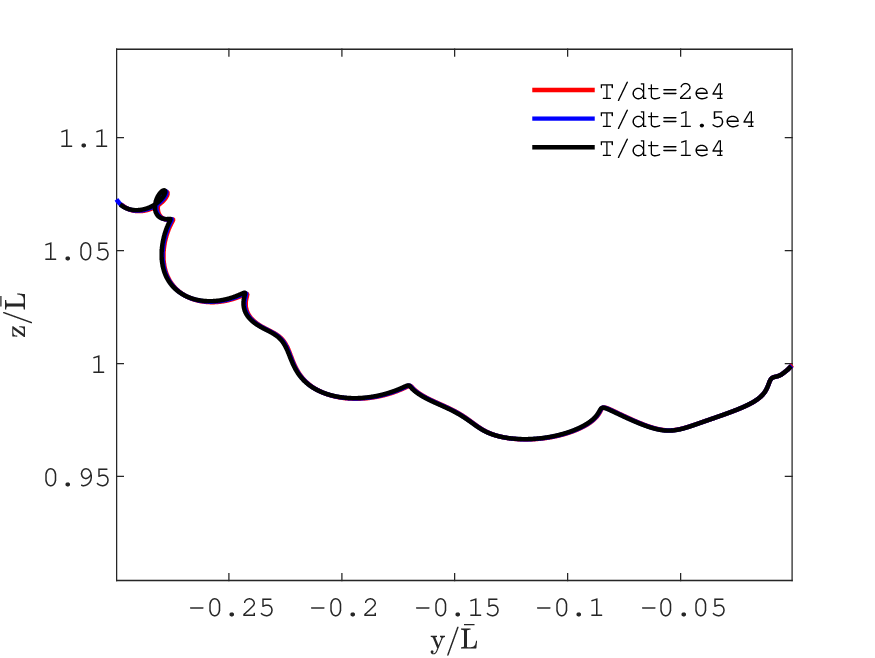}
         \caption{}
     \end{subfigure}
     \hfill
     \begin{subfigure}{0.49\textwidth}
         \centering
         \includegraphics[scale=0.56]{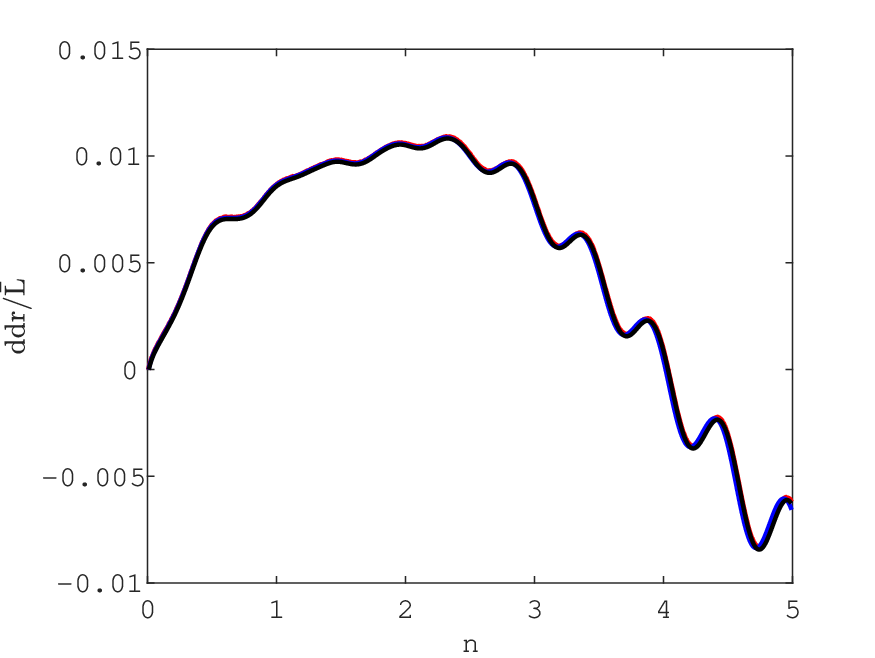}
         \caption{}
     \end{subfigure}
     \hfill
        \caption{(a) Phase plot and (b) variation of lateral drift with an increase in the number of swimming cycles, n. Numerical convergence test of 2 swimmers for different time steps, dt and a mesh size ds=L/1x10$^4$.}
        \label{fig:numericsdt}
\end{figure*}

\begin{figure*}
     \centering
     \begin{subfigure}{0.49\textwidth}
         \centering
         \includegraphics[scale=0.56]{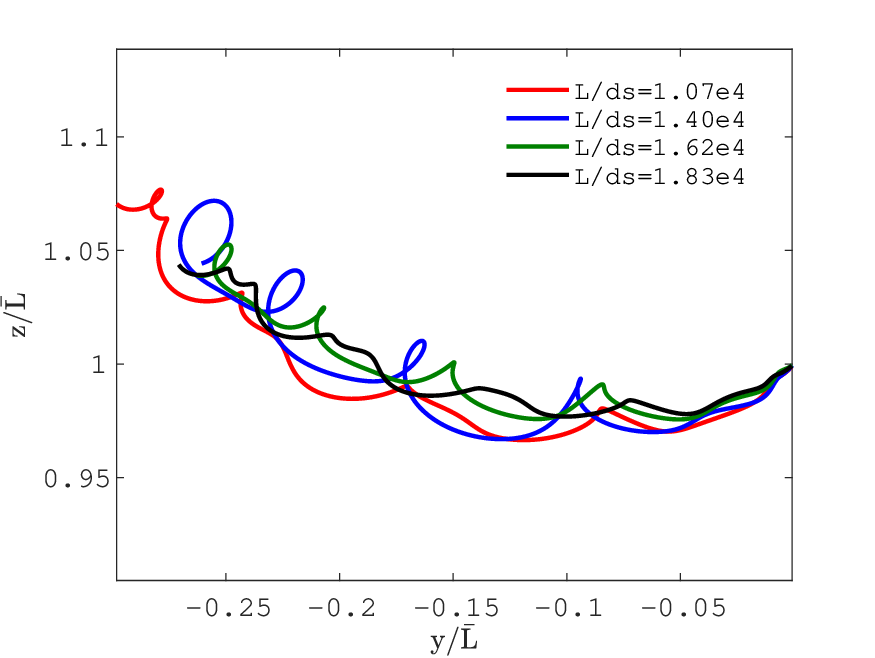}
         \caption{}
     \end{subfigure}
     \hfill
     \begin{subfigure}{0.49\textwidth}
         \centering
         \includegraphics[scale=0.56]{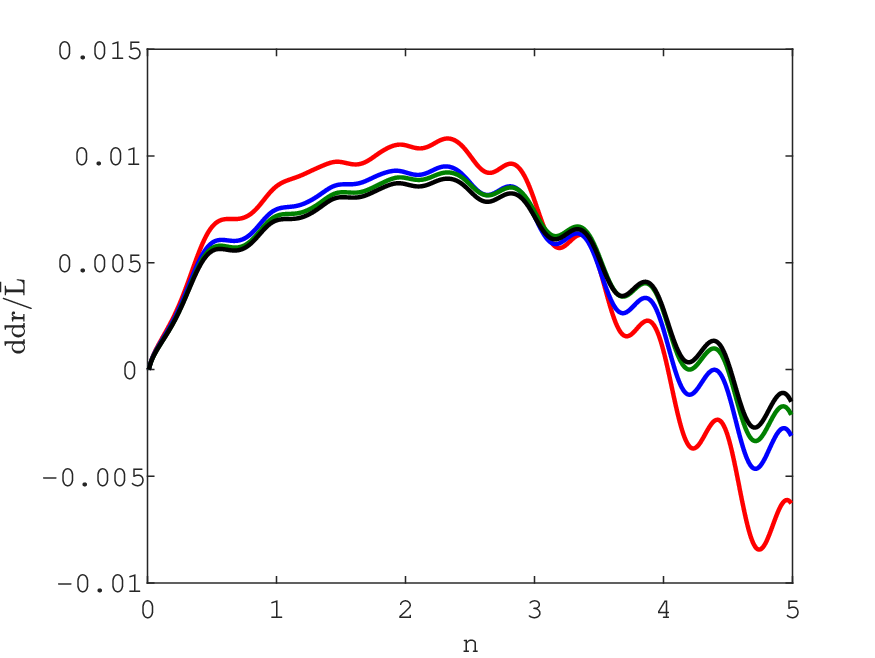}
         \caption{}
     \end{subfigure}
     \hfill
        \caption{(a) Phase plot and (b) variation of lateral drift with an increase in the number of swimming cycles, n. Numerical convergence test of 2 swimmers for different mesh sizes, ds and a time step dt=T/1x10$^4$.}
        \label{fig:numericsds}
\end{figure*}

We report their swimming trajectories (phase plots) and the normalized lateral separation (drift dr/$\mathrm{\Bar{L}}$) in the y-z plane (to completely understand and quantify the hydrodynamic interactions between them). To further validate our numerical framework, we study this example for different values of dt (see Fig. \ref{fig:numericsdt}) and ds (see Fig. \ref{fig:numericsds}). Note that we plot the phase plots only for swimmer s1 (for the sake of clarity). Owing to symmetry, the trajectory of swimmer s2 would be alike. We observe that the choice of different time steps does not have a huge influence on either the phase plot (Fig. \ref{fig:numericsdt}a) or the lateral drift (Fig. \ref{fig:numericsdt}b). Two length scales appear in the phase plot; while we attribute the smaller length scale to the typical wiggling motion of the geometric center of the helical swimmer itself, the larger length scale represents the circular path traversal of the swimmers, see Fig. \ref{fig:numericsds}b. These two length scales could interact in order to generate a third length scale (as we shall later see for robots of different aspect ratios).

\begin{figure*}
     \centering
     \begin{subfigure}{0.49\textwidth}
         \centering
         \includegraphics[scale=0.56]{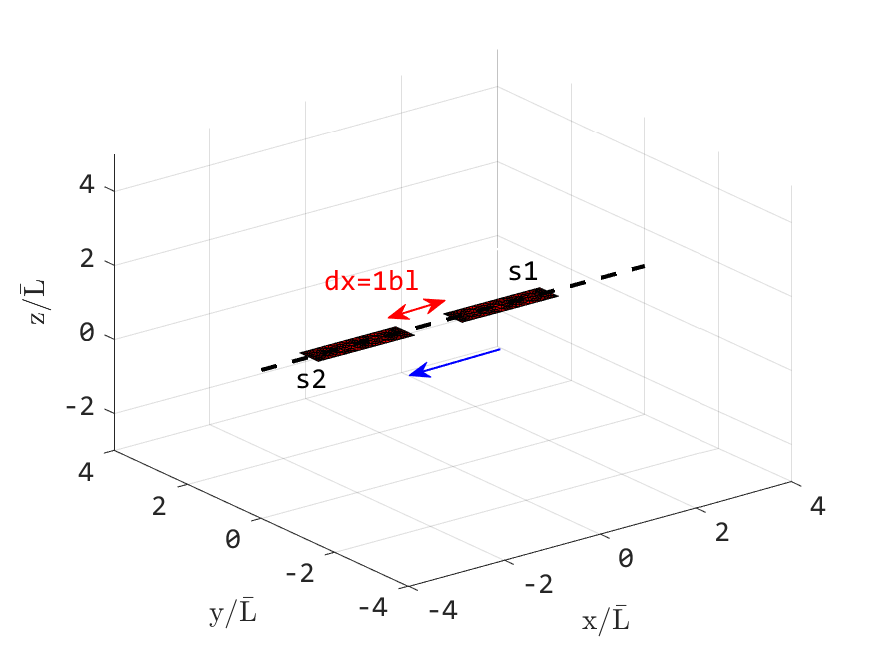}
         \caption{dx=1bl, dz=0bl.}
     \end{subfigure}
     \hfill
     \begin{subfigure}{0.49\textwidth}
         \centering
         \includegraphics[scale=0.56]{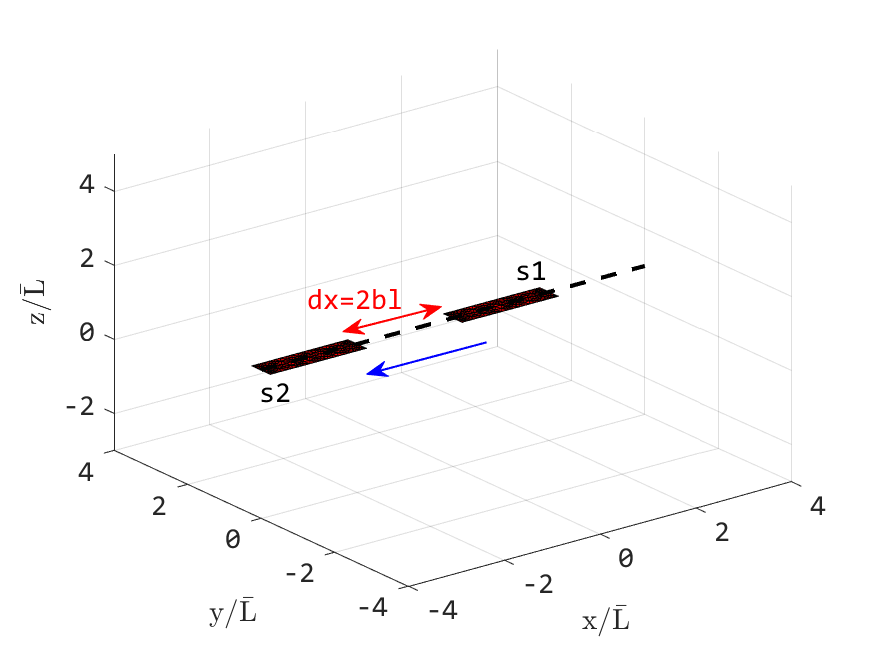}
         \caption{dx=2bl, dz=0bl.}
     \end{subfigure}
     \hfill
     \begin{subfigure}{0.49\textwidth}
         \centering
         \includegraphics[scale=0.56]{Figures/snapshot2sdx0bldz2bl.eps}
         \caption{dx=0bl, dz=2bl.}
     \end{subfigure}
     \hfill
     \begin{subfigure}{0.49\textwidth}
         \centering
         \includegraphics[scale=0.56]{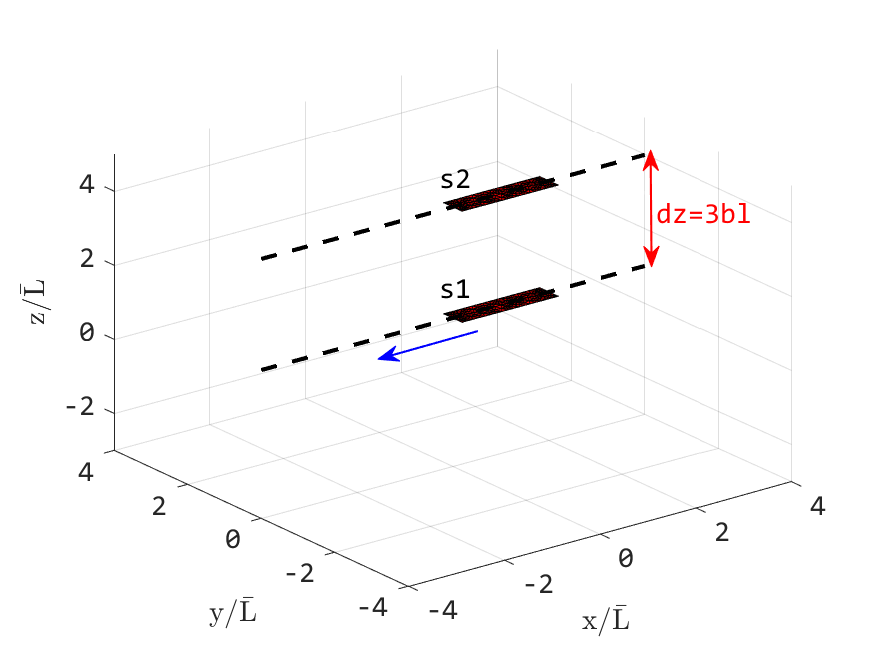}
         \caption{dx=0bl, dz=3bl.}
     \end{subfigure}
     \hfill
        \caption{Schematic representation of different initial (spatial) configurations of two soft robotic swimmers. Swimmers s1 and s2 are identified in black. The direction of swimming is shown in blue. The separation between the swimmers is denoted as either dx or dz in red. The front of swimmer s1 in all subplots is located at (x,y,z) = (0,0,$\mathrm{\Bar{L}}$); for (a) and (c): (multimedia available online).}
        \label{fig:snapshots2s}
\end{figure*}

\subsection{Emergent behavior: 2 swimmers}

Now, we systematically consider different in-plane and out-of-plane configurations to study the effect of initial arrangement (spatial orientation) upon their collective swimming, spatiotemporal patterning, and emergent behavior. First, we study the 2-swimmer swarm that are spaced a few bodylengths (bl) apart along either the x- or z-axis (see Fig. \ref{fig:snapshots2s}). Due to location motion direction into the x-axis and the angular motion in the y-z plane, a spacing in the y-direction is similar to a spacing in the z-direction. We define dr/$\mathrm{\Bar{L}}$ as the normalized lateral separation between the swimmers, while ddr/$\mathrm{\Bar{L}}$ measures the change in dr/$\mathrm{\Bar{L}}$ to represent the lateral drift. Furthermore, we have similar notations for axial separation: dx/$\mathrm{\Bar{L}}$, and ddx/$\mathrm{\Bar{L}}$ represents the normalized axial separation and change in axial separation between the swimmers, and so on. Note that we report non-dimensional kinematic quantities such as phase plots, axial approach (ddx/$\mathrm{\Bar{L}}$) and lateral drift (ddr/$\mathrm{\Bar{L}}$) as a function of number of actuation cycles (n) to ensure a uniform comparison of their swimming trajectories and spatiotemporal patterning.

\begin{figure*}
     \centering
     \begin{subfigure}{0.49\textwidth}
         \centering
         \includegraphics[scale=0.56]{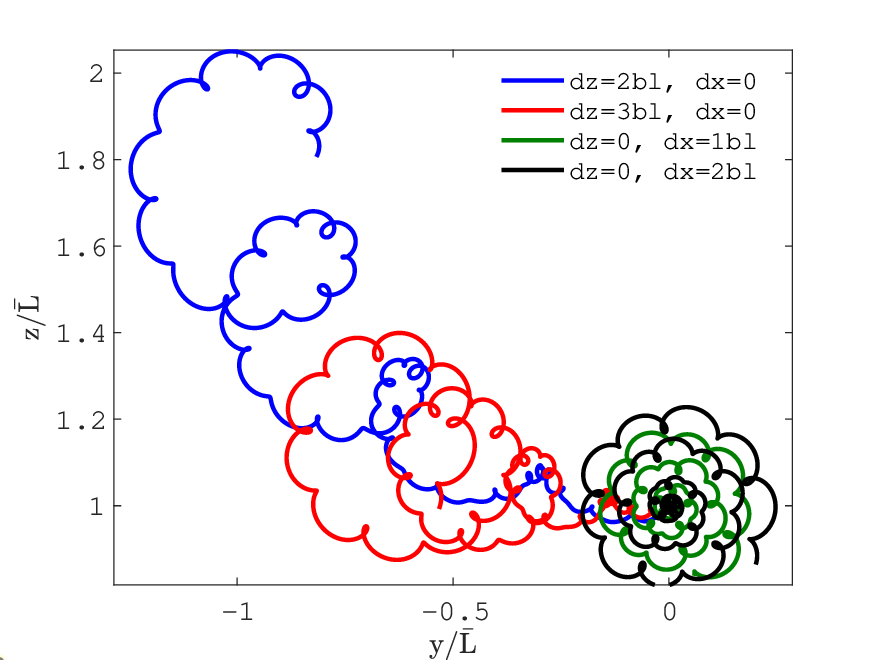}
         \caption{}
     \end{subfigure}
     \hfill
     \begin{subfigure}{0.49\textwidth}
         \centering
         \includegraphics[scale=0.56]{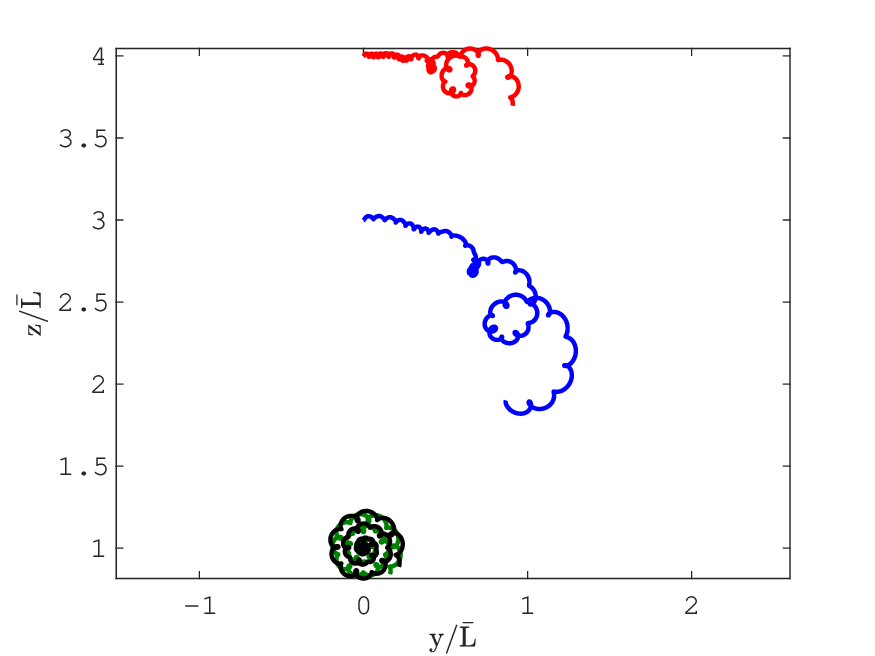}
         \caption{}
     \end{subfigure}
     \hfill
     \begin{subfigure}{0.49\textwidth}
         \centering
         \includegraphics[scale=0.56]{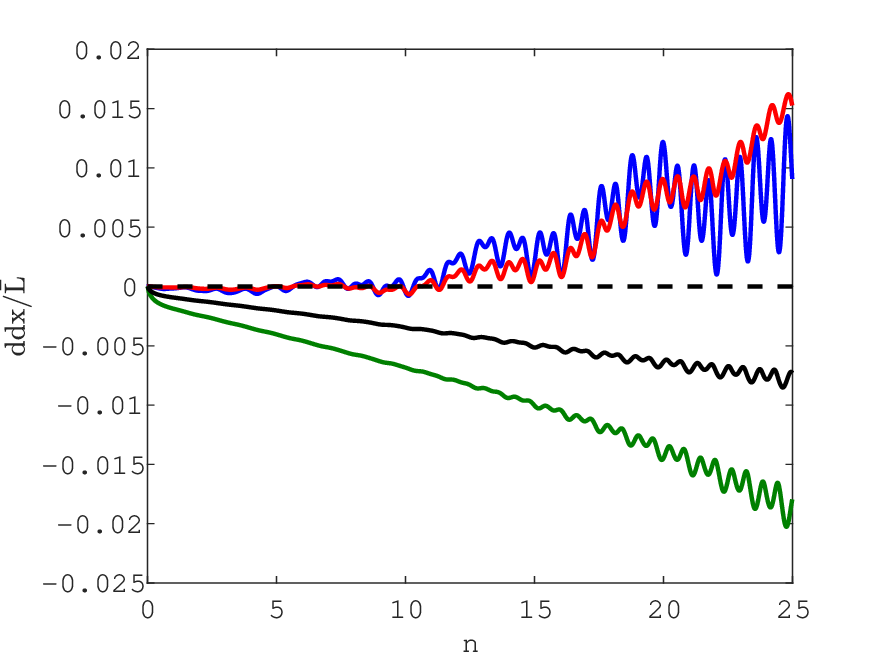}
         \caption{}
     \end{subfigure}
     \hfill
     \begin{subfigure}{0.49\textwidth}
         \centering
         \includegraphics[scale=0.56]{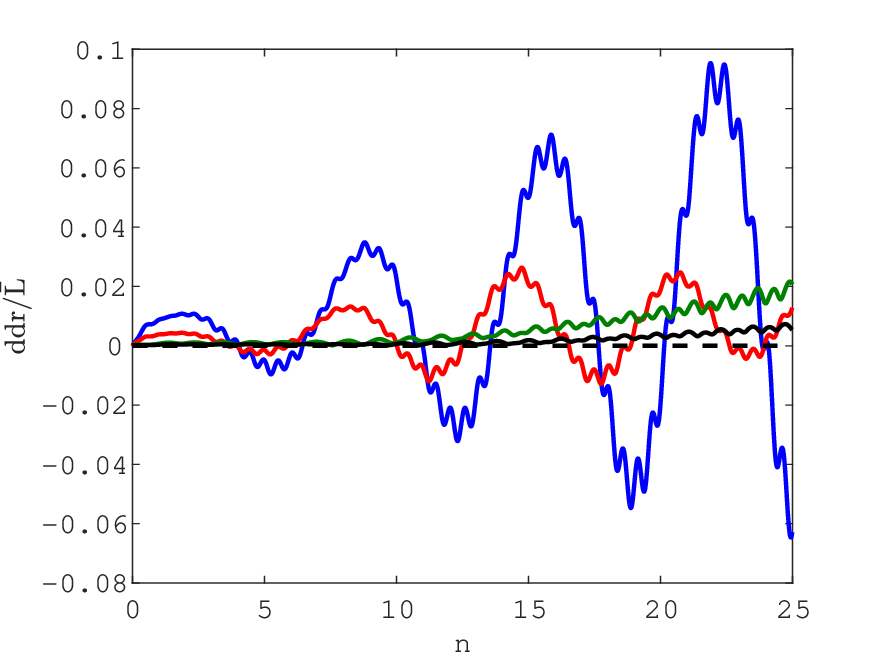}
         \caption{}
     \end{subfigure}
     \hfill
        \caption{(a) Swimming trajectories of (a) s1 and (b) s2, and variation of normalized (c) longitudinal and (d) lateral separation over each swimming cycle. Note that the swimmers gradually drift laterally outward in a sinusoidal pattern. This is apparent from the curve's bias, which shows an increasing net area as the number of swimming cycles increases.}
        \label{fig:2s}
\end{figure*}

Four different starting configurations are considered: (a) dx=1bl, dz=0bl [Fig. \ref{fig:snapshots2s}a (multimedia online)], (b) dx=2bl, dz=0bl (Fig. \ref{fig:snapshots2s}b), (c) dx=0bl, dz=2bl [Fig. \ref{fig:snapshots2s}c (multimedia online)], and (d) dx=0bl, dz=3bl (Fig. \ref{fig:snapshots2s}d). While (a) and (b) are representative of the axial (longitudinal) separation, (c) and (d) account for the lateral separation. The results are presented in Fig. \ref{fig:2s}. Note that the hydrodynamic effects become more pronounced when the swimmers are in the preferential plane (here, y-z plane) due to the typical nature of the flow field generated by these swimmers (multimedia available online). Consequently, the lateral outward drift is more pronounced for (c) and (d), although there is also an outward drift for (a) and (b), see Fig. \ref{fig:2s}d. The configurations (a) and (b) are quite similar and so are the hydrodynamic interactions between them. In fact, they differ only by magnitude, where (a) can be perceived as a future event of (b), when the axial approach brings these swimmers close enough to each other by 1bl. Similarly, the configurations (c) and (d) are also very similar and differ only by magnitude, where (d) can potentially be perceived as a future event of (c) when the lateral outward drift sets these swimmers apart further by a spacing of an additional 1bl.

As discussed previously, the swimming trajectory of a swimmer is quite similar to a fractal (geometric) structure. Taking a close look at the phase plot of swimmer s1 (blue curve in Fig. \ref{fig:2s}a), we observe three different length scales. The wiggling (helical) motion of the soft robotic swimmer with every actuation cycle is manifested in the smallest length scale (a half-circle). The largest length scale is the outcome of hydrodynamic interactions between the individual swimmers when they revolve around each other. The third length scale (cloud-like structure) is in between the other two extreme length scales.

\begin{figure*}
     \centering
     \begin{subfigure}{0.49\textwidth}
         \centering
         \includegraphics[scale=0.56]{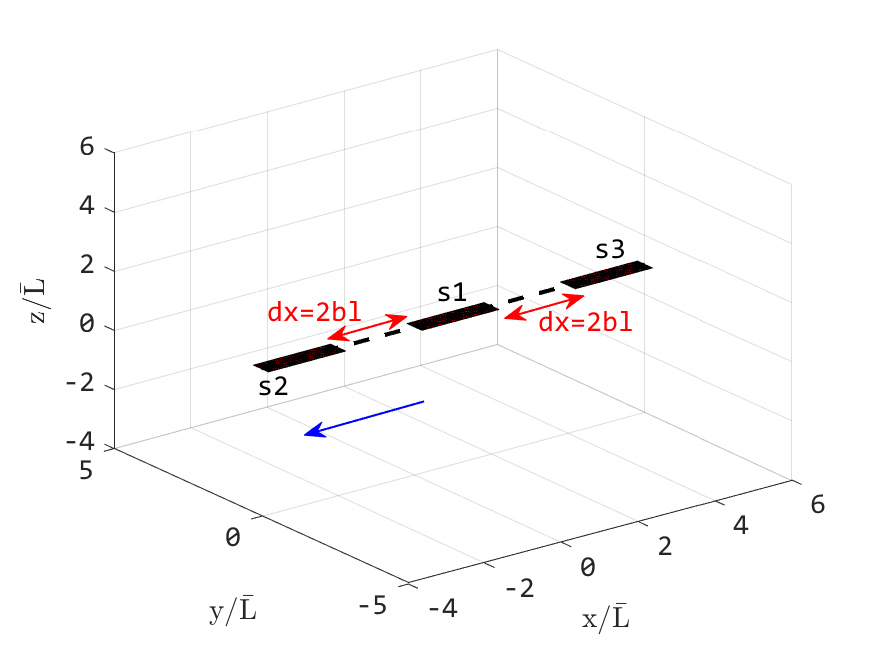}
         \caption{dx=2bl, dz=0bl.}
     \end{subfigure}
     \hfill
     \begin{subfigure}{0.49\textwidth}
         \centering
         \includegraphics[scale=0.56]{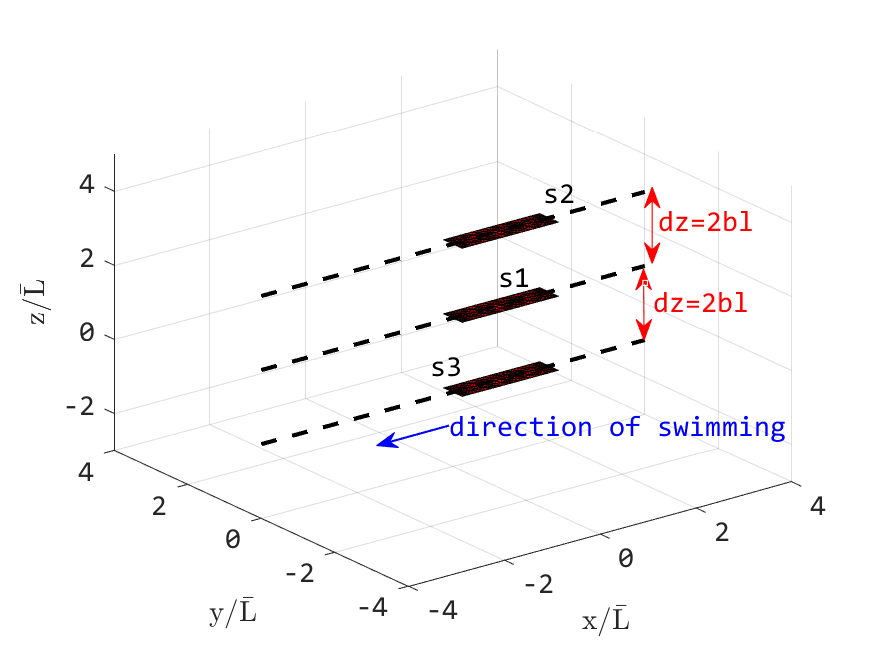}
         \caption{dx=0bl, dz=2bl.}
     \end{subfigure}
     \hfill
     \begin{subfigure}{0.49\textwidth}
         \centering
         \includegraphics[scale=0.56]{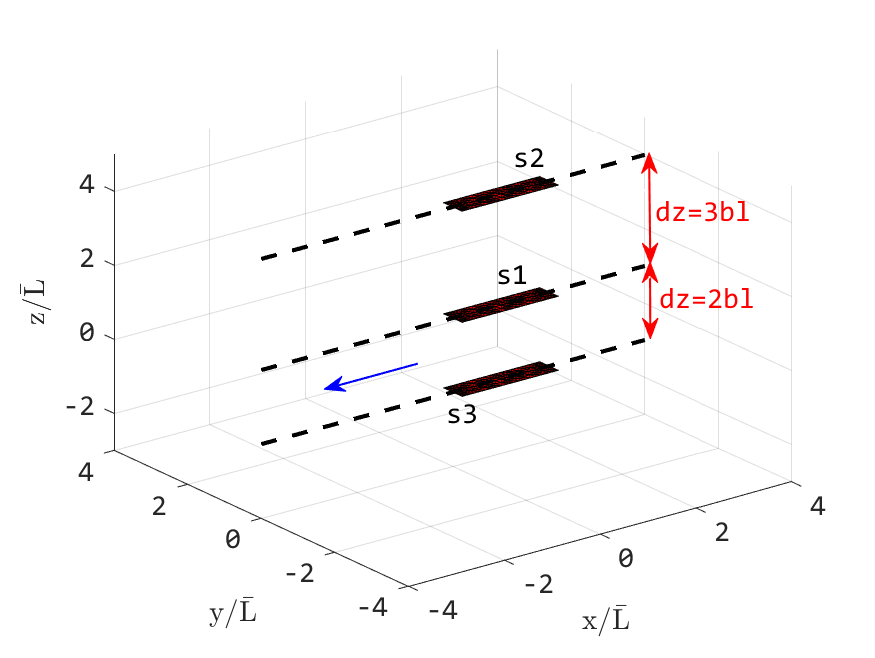}
         \caption{dx=0bl, dz13=2bl, dz12=3bl.}
     \end{subfigure}
     \hfill
     \begin{subfigure}{0.49\textwidth}
         \centering
         \includegraphics[scale=0.56]{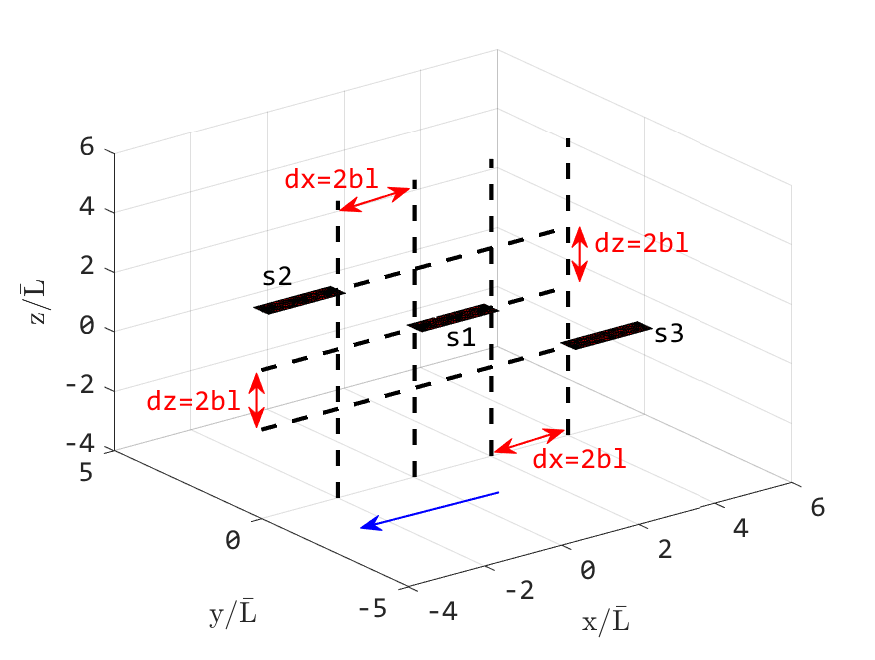}
         \caption{dx=2bl, dz=2bl.}
     \end{subfigure}
     \hfill
        \caption{Schematic representation of different initial (spatial) configurations of three soft robotic swimmers. Swimmers s1, s2, and s3 are identified in black. The direction of swimming is shown in blue. The separation between the swimmers is denoted (dx and/or dz) in red. The front of swimmer s1 in all subplots is located at (x,y,z) = (0,0,$\mathrm{\Bar{L}}$); for (a) and (b): (multimedia available online).}
        \label{fig:snapshots3s}
\end{figure*}

While the swimming trajectory of the swimmer s1 for the configuration in Fig. \ref{fig:snapshots2s}d is similar to that for the configuration in Fig. \ref{fig:snapshots2s}c, the hydrodynamic interactions and the rotational rearrangement are higher in magnitude (see red and blue curves in Fig. \ref{fig:2s}a). Therefore, all the three length scales grow in magnitude as does ddr/$\mathrm{\Bar{L}}$ in Fig. \ref{fig:2s}d, although they still exhibit a fractal structure in their swimming trajectories. The swimming trajectories of swimmer s2 for the aforementioned four distinct starting configurations are schematically plotted in Fig. \ref{fig:2s}b. Clearly, the blue trajectory is the outcome when they are place 2bl apart along the lateral direction, while the red trajectory represents the phase plot of swimmer s2 when their (initial) separation is 3bl.

When the swimmers have an axial separation (and no lateral separation), the hydrodynamic interactions seem to be the least, although they still influence the swimming kinematics of each other. Therefore, the swimmers still exhibit the smallest length scale, although the other two length scales are insignificant (see Figs. \ref{fig:2s}a and \ref{fig:2s}b). However, there is important spatiotemporal manifestations even during these cases. It turns out they will approach each other which might be caused by the front swimmer causing a reduction of drag forces for the follower swimmer. Therefore, this helps the latter to swim faster and move closer to the leading swimmer with every swimming cycle. Hence, they behave much like attractors, wherein the normalized distance between them gradually decreases with the number of actuation cycles (as shown in Fig. \ref{fig:2s}c).

\begin{figure*}
     \centering
     \begin{subfigure}{0.49\textwidth}
         \centering
         \includegraphics[scale=0.56]{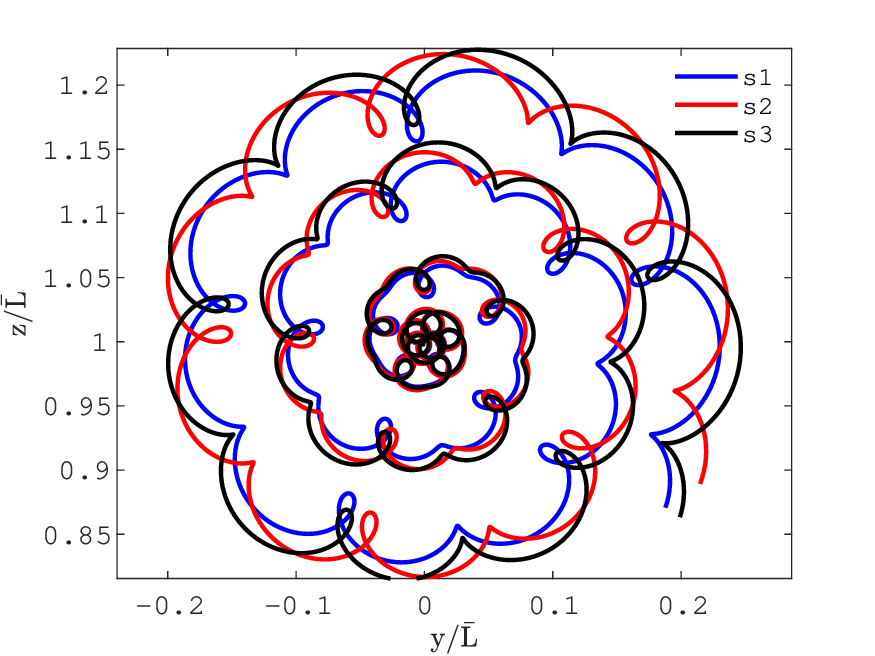}
         \caption{}
     \end{subfigure}
     \hfill
     \begin{subfigure}{0.49\textwidth}
         \centering
         \includegraphics[scale=0.56]{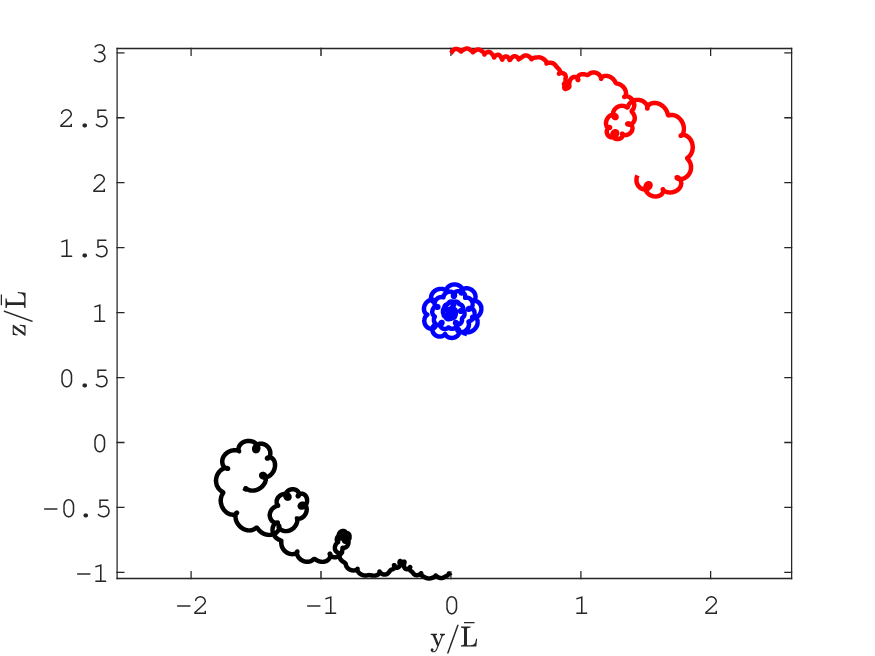}
         \caption{}
     \end{subfigure}
     \hfill
     \begin{subfigure}{0.49\textwidth}
         \centering
         \includegraphics[scale=0.56]{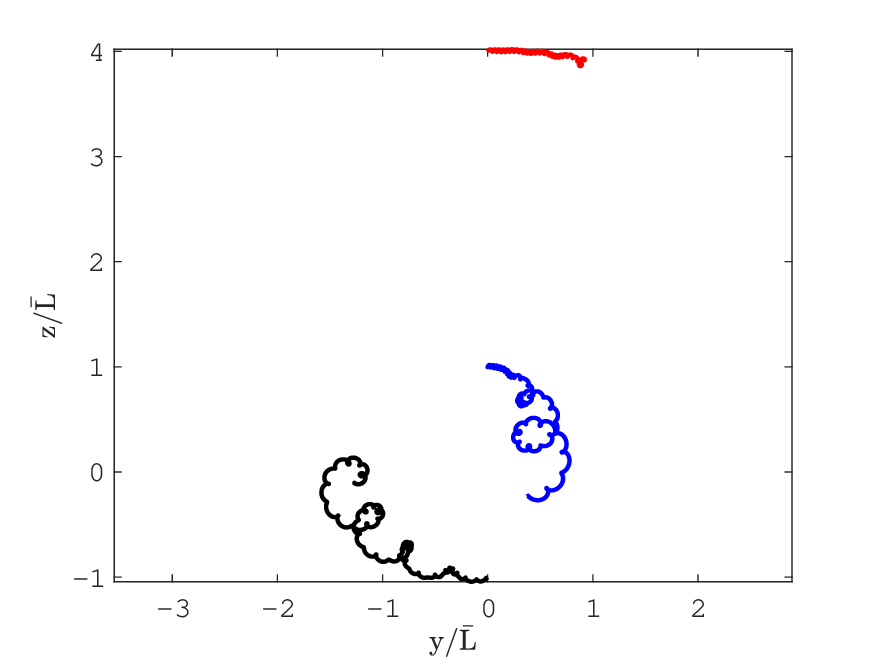}
         \caption{}
     \end{subfigure}
     \hfill
     \begin{subfigure}{0.49\textwidth}
         \centering
         \includegraphics[scale=0.56]{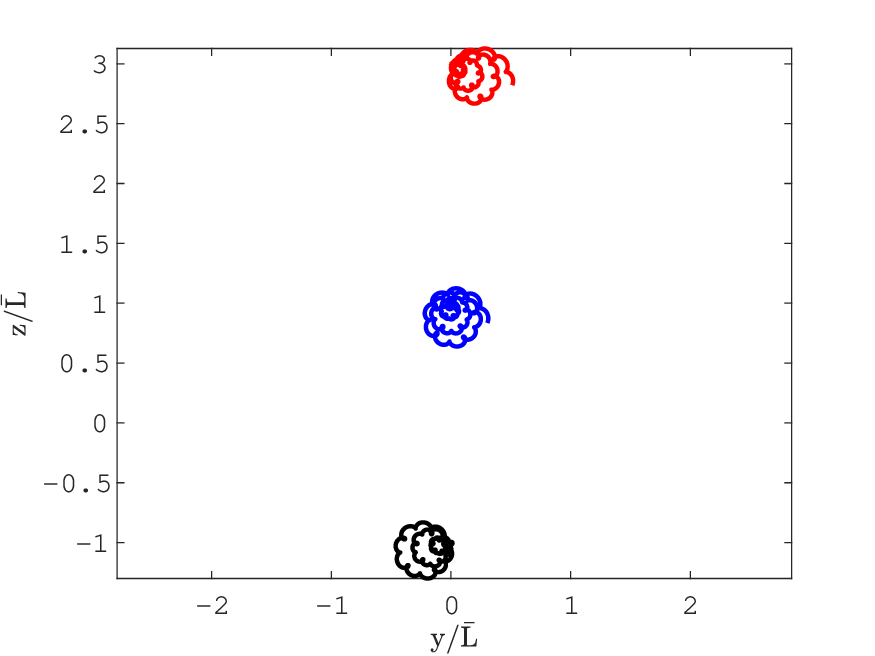}
         \caption{}
     \end{subfigure}
     \hfill
        \caption{Swimming trajectories of three swimmers for different spatial configurations in (a), (b), (c), and (d), corresponding to the starting configurations shown in Figs. \ref{fig:snapshots3s}(a), \ref{fig:snapshots3s}(b), \ref{fig:snapshots3s}(c), and \ref{fig:snapshots3s}(d), respectively (multimedia available online).}
        \label{fig:3s}
\end{figure*}

\begin{figure}[htpb!]
    \centering
    \includegraphics[scale=0.56]{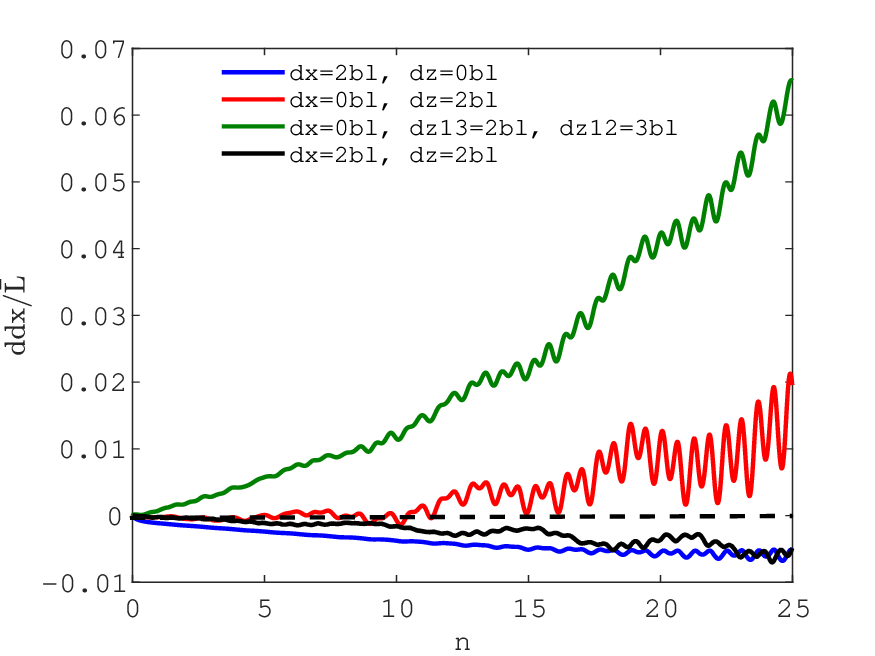}
    \caption{Variation of normalized change in axial separation between swimmers s2 and s3 over each swimming cycle for different configurations of three swimmers depicted in Fig. \ref{fig:snapshots3s}.}
    \label{fig:3sdx}
\end{figure}

It is important to note that as they come closer to each other, they also start interacting differently because they now see each other in a more pronounced manner causing them to start moving laterally outward (although quite minimally; see Fig. \ref{fig:2s}d), and this is clearly manifested in the increasing oscillatory pattern after n=15 in Figs. \ref{fig:2s}c and \ref{fig:2s}d. However, when there is a lateral separation (and no axial separation) between the swimmers, the hydrodynamic interactions are most prominent. Therefore, the individual swimmers revolve around each other and have the most circular distance covered during their swimming trajectory. Typically, they behave as repellers and the normalized lateral distance between them sinusoidally varies (with a time-average increase) with an increasing number of cycles (see Fig. \ref{fig:2s}d). Importantly, the cumulative area under all the curves in \ref{fig:2s}d is positive (irrespective of whether dr/$\mathrm{\Bar{L}}$ varies sinusoidally or increases almost linearly with small oscillations). This means that the swimmers drift apart with every actuation cycle.

Briefly, after analyzing the fluid-mediated hydrodynamic interactions between two soft robotic swimmers and also studying the influence of different starting configurations, the following picture emerges: the out-of-plane swimmers come closer to each other (attractors) and their axial separation decreases; essentially, the hydrodynamic interactions gradually increase, until they can potentially lie in one single (preferential plane). Once in the in-plane configuration, the swimmers start drifting laterally outwards to reduce the hydrodynamic interactions (repellers) until they do not see each other anymore (theoretically, at infinity). This gradual transition from an out-of-plane configuration to slowly merging to a preferred in-plane configuration seems to be the hallmark of emergent behavior of swarms of helical soft robotic swimmers. Although we have, until now, shown this (emergent) behavior only for two swimmers, this hypothesis likely holds true for larger systems (e.g., swarms).

\subsection{Emergent behavior: 3 swimmers}
Next, we study the spatiotemporal patterning and emergent behavior of three swimmers. Like before, we consider three similar helical soft robotic swimmers (s1, s2, and s3) for four distinct starting configurations: (a) when they are axially separated by 2bl [Fig. \ref{fig:snapshots3s}a (multimedia online)], (b) when they are laterally separated by 2bl [Fig. \ref{fig:snapshots3s}b (multimedia online)], (c) when they are laterally separated but asymmetrically by 2bl and 3bl (Fig. \ref{fig:snapshots3s}c), and (d) when they are diagonally arranged as shown in Fig. \ref{fig:snapshots3s}d; the phase plots of these four cases for n=25 swimming cycles are plotted in Figs. \ref{fig:3s}a-\ref{fig:3s}d, respectively.

When the swimmers are axially separated with no lateral separation initially, the hydrodynamic interactions between them are minimal. This is intuitive from our previous analysis of the two swimmer swarms. Therefore, their phase plots (swimming trajectories) look quite similar to each other (see Fig. \ref{fig:3s}a). Furthermore, thanks to negligible fluidic cross-talk between the individual swimmers in this configuration, there exists only one predominant length scale (i.e., the smallest wiggling helical motion manifested length scale). The fractal structure is missing, and so is the largest length scale of circular path traversal, because the centers of rotation of all the three swimmers are collinear in the axially separated configuration at (y,z)=(0,$\mathrm{\Bar{L}}$). Nevertheless, this configuration (if continued for several hundreds of actuation or swimming cycles) would gradually result in an in-plane (i.e., preferential plane) configuration - maintaining the hallmark emergent behavior of these helical soft robotic swimmer collectives (see Fig. \ref{fig:3sdx}). Finally, upon a close look into the swimming trajectories, we observe a laterally outward trajectory, much like a spiral curve with increasing radius (reminiscent of the famous Fermat's spiral).

Furthermore, there appears to be an increasing phase lag between the red (s2) and the blue/black swimmer (s1/s3). Our reasoning for this observation again goes back to our previous study on two axially separated swimmer. Here, swimmer s2 is the leading swimmer, and faces the maximum fluidic resistance with the trailing swimmers s1 and s3 having lower drag forces. Hence, s1 and s3 approach s2 aiming to minimize their axial separation. Hence, there is a phase lag in the swimming trajectory of swimmer s2 (in red) compared to that of swimmers s1 (in blue) or s3 (in black), both of which are similar without noticeable phase lag.

In Fig. \ref{fig:3s}b, we observe the phase plot of laterally separated symmetric in-plane three swimmer swarms. This is typically the situation when the hydrodynamic interactions are maximal. Consequently, the outer swimmers s2 and s3 start revolving around the central swimmer (s1) and drift radially outward. Owing to the symmetry of the configuration, the phase plot of the outer swimmers are quite similar. Additionally, they exhibit a fractal structure as discussed previously. The outer swimmers eventually drift apart laterally outwards until they negligibly interact any further (theoretically, at infinity). However, they still have a sinusoidally varying lateral distance between them (that has a positive area under the curve) with every swimming/actuation cycle. While the swimmer s2 (in red) travels initially rightward, the swimmer s3 (in black) traverses leftward in the same way due to the nature of the flow field around these swimmers (multimedia available online).

The configuration of the soft robotic swimmers in Fig. \ref{fig:3s}c is similar to Fig. \ref{fig:3s}b, apart from the fact that the outer swimmers are not symmetrically placed initially. Therefore, all the three swimmers start revolving along their distinct (fractal-like) swimming paths (in spiral trajectories) radially outward. All the swimmers revolve clockwise in line with the surrounding flow field (multimedia available online). The hydrodynamic effects on the swimming trajectory of swimmer s2 are the least. Consequently, it revolves much less and has traversed a minimal path of revolution.

Finally, we investigate the configuration as schematically shown in Fig. \ref{fig:snapshots3s}d. These swimmers are initially placed in a diagonal manner. Here, it would be interesting to investigate whether their motion can be explained from a superposition of the axial motion of the swimmers (in case they do not have a lateral separation (Fig. \ref{fig:snapshots3s}a) and the lateral motion (in case they do not have an axial separation, as in \ref{fig:snapshots3s}b). Therefore, we plot their axial behaviour in Fig. \ref{fig:3sdx}. However, we notice that this is not the case. In both the axial and lateral direction they interact much less than in the case they would only have axial or lateral separation. Fig. \ref{fig:3s}d is not the same as Fig. \ref{fig:3s}b. In addition, their axial motion is not the same as that of the case of Fig. \ref{fig:snapshots3s}a. This is simply because of their reduced hydrodynamic interactions owing to their initial distance being larger than in the case of only lateral or axial separation.

\subsection{Influence of aspect ratio}
\begin{figure}[htpb!]
    \centering
    \includegraphics[scale=0.25]{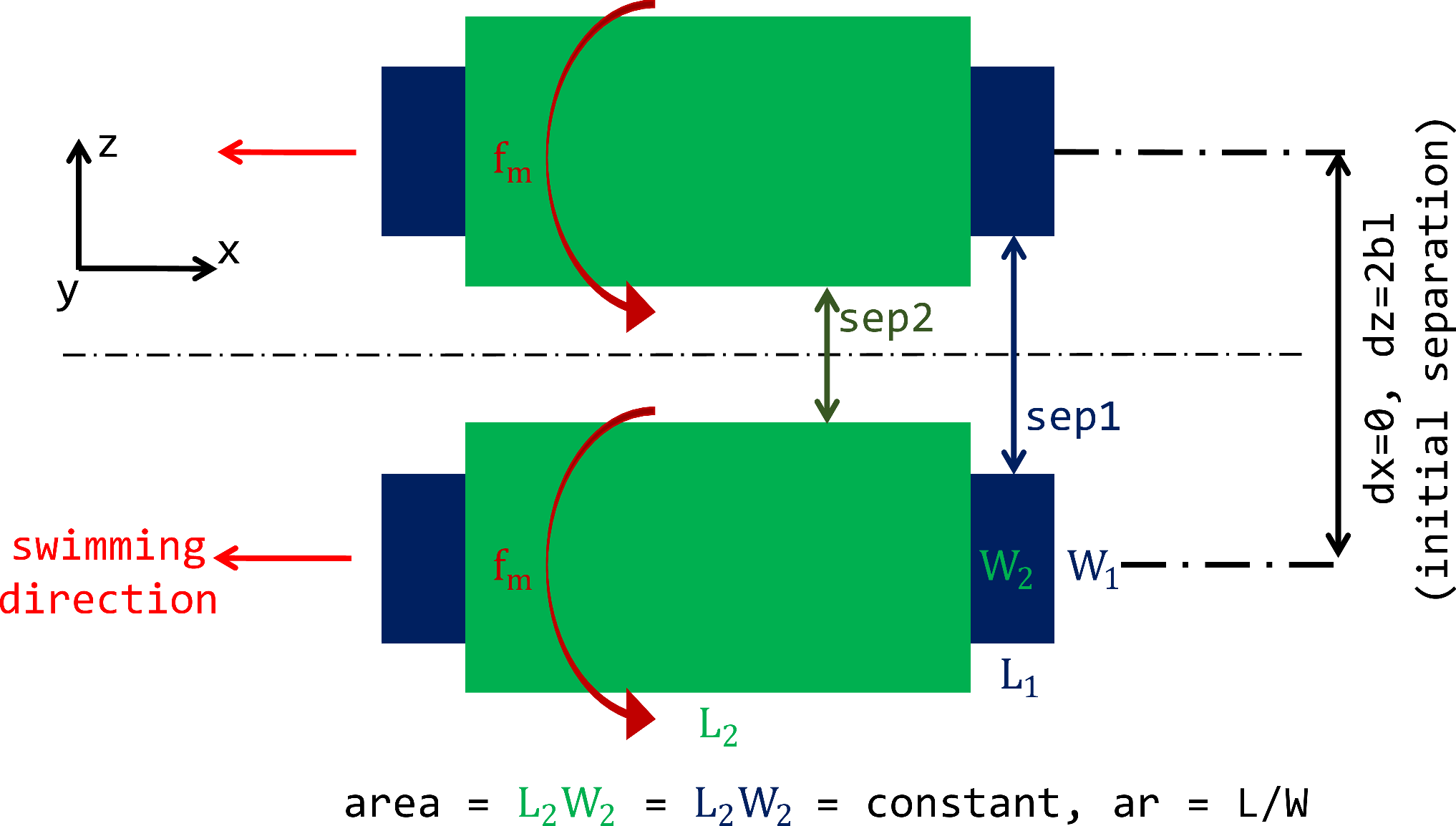}
    \caption{Schematic representation of swimmers with different aspect ratios, yet having the same area. The swimmers have an initial lateral separation of dz=2bl, and axial separation of dx=0 irrespective of their aspect ratios. However, a higher ar (blue swimmer) has a larger lateral separation (sep1), and thus account for lower hydrodynamic interactions. The swimmers with a lower ar see each other more (due to lower value of sep2), and therefore, interact more through fluidic cross-talk. Please note that the difference between sep1 and sep2 is minimal compared to dz=2bl; i.e., sep1-sep2$\ll$2bl. Figure is for representation only, and not to scale.}
    \label{fig:ar}
\end{figure}

\begin{figure*}
     \centering
     \begin{subfigure}{0.49\textwidth}
         \centering
         \includegraphics[scale=0.56]{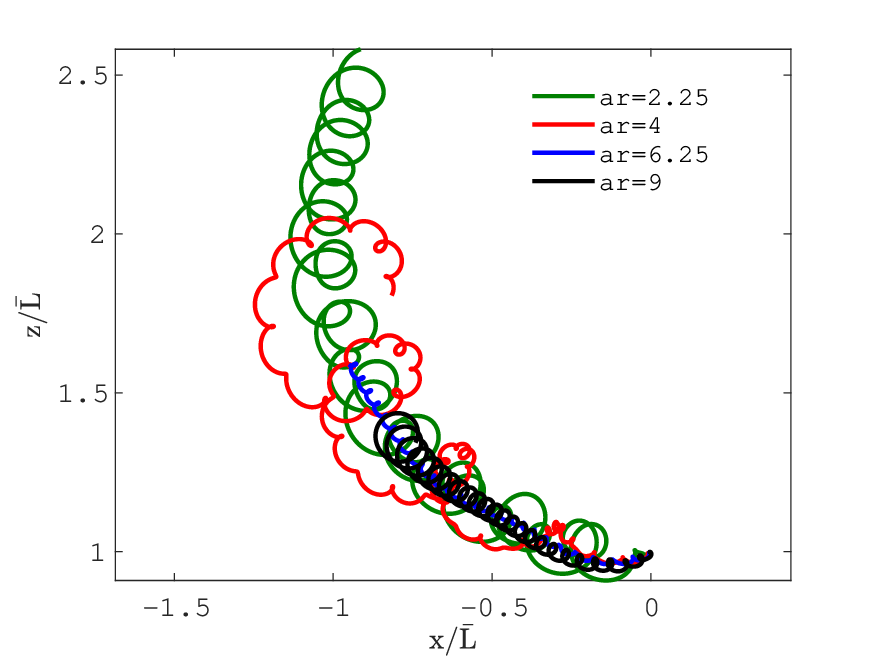}
         \caption{}
     \end{subfigure}
     \hfill
     \begin{subfigure}{0.49\textwidth}
         \centering
         \includegraphics[scale=0.56]{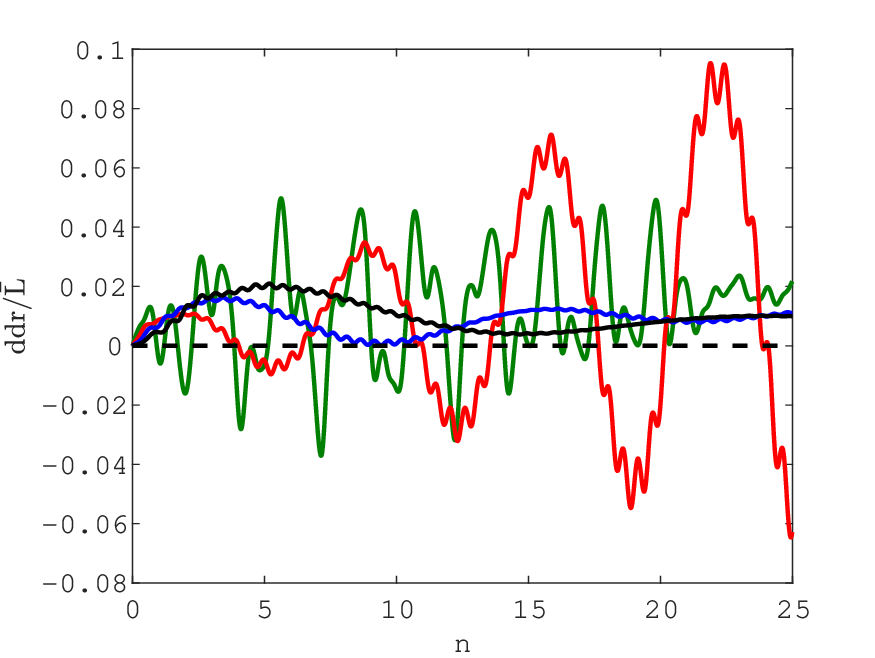}
         \caption{}
     \end{subfigure}
     \hfill
        \caption{(a) Phase plots and (b) variation of normalized lateral separation with number of swimming cycles of 2 swimmers for different aspect ratios.}
        \label{fig:2sdz10aspectratio}
\end{figure*}

Finally, we investigate the role of aspect ratio (ar) upon the hydrodynamic interactions between the soft robotic swimmers, which result in slightly different emergent behavior, phase plots, and spatiotemporal patterning. Here, the swimmer area (LW) is kept the same (for a uniform analysis), while ar is changed simply by varying the length L and width W systematically (see Fig. \ref{fig:ar}); and, ar = L/W. Furthermore, the characteristic body length $\mathrm{\Bar{L}}$ is fixed by virtue of its definition, and is the square root of the swimmer area (see Fig. \ref{fig:helicalswimmer}). To investigate this further, we consider the configuration of two swimmers that are laterally separated by dz=2bl (see Figs. \ref{fig:snapshotnumerics}, \ref{fig:snapshots2s} and \ref{fig:2s}, having an ar of 4). The phase plot of swimmer s1 is represented in Fig. \ref{fig:2sdz10aspectratio}a, where we note that the case of ar=4 corresponds to the results shown in Fig. \ref{fig:2s}a. The normalized lateral separation between them as a function of number of cycles, n is plotted in Fig. \ref{fig:2sdz10aspectratio}b for different values of ar ranging from 2.25 to 9, where the swimmers with the ar value of 2.25 is less rectangular (and thus difficult to twist), while the ones with ar=9 can easily twist.

Hence, for the same initial spatial configuration (dx=0bl, dz=2bl), the swimmers with a higher aspect ratio (but same area, and therefore, lower width \textit{W}) has a higher lateral separation, effectively leading to lower hydrodynamic interactions (see Fig. \ref{fig:ar}). Therefore, the swimmers of ar=9 have a slightly higher separation (compared to swimmers of ar=2.25) and they interact less through fluidic cross-talk). The phase plot of swimmer s1 for ar=2.25 has the highest circular distance covered for the same number of swimming cycles (n=25), compared to that of swimmer s1 for ar=9 that has the lowest distance (see Fig. \ref{fig:2sdz10aspectratio}a).

\section{\label{sec:level3}CONCLUSION}

In conclusion, we report the emergent behavior, collective swimming, and spatiotemporal patterning of magnetically actuated miniaturized helical soft robotic swarms consisting of 2 to 3 swimmers, where typically, the individual swimmers affect the swimming behavior of each other through long-range fluid-mediated non-reciprocal hydrodynamic interactions leading to far-from-equilibrium self-organizational behavior. When these swimmers are out of plane (i.e., they have a longitudinal separation), they gradually come closer to each other over each swimming cycle, thereby behaving as attractors. However, when they are in plane (i.e., no longitudinal separation, but only a lateral separation), they gradually move outward along a lateral direction (thereby behaving as repellers). Therefore, they evade any possibility of agglomeration (or clustering). Although we report our findings mostly for two and three swimmer configurations, our results can principally be extrapolated to larger systems (i.e., swarms) without any loss of generality.

\section*{ACKNOWLEDGEMENTS}
The authors would like to thank the Center for Information Technology of the University of Groningen for their support and for providing access to the Habrok high-performance computing cluster. The work is carried out within the research program of the Centre for Data Science and Systems Complexity (DSSC), Faculty of Science and Engineering, University of Groningen.


\section*{Conflict of interest}
The authors have no conflicts to disclose.



\bibliography{apssamp}

\end{document}